%% file: main.tex
\documentclass[%
amsmath,aps,showkeys,
amssymb,twocolumn,prx,
superscriptaddress
]{revtex4-2}

\usepackage{hyperref}
\hypersetup{
    colorlinks,
    linkcolor={black},
    citecolor={blue!80!black},
    urlcolor={blue!80!black}
}

\usepackage{graphicx}
\usepackage{bm}
\usepackage{bbold}
\usepackage{dcolumn}
\newcolumntype{d}[1]{D{.}{.}{#1}}

\usepackage[dvipsnames]{xcolor}
\newcommand{\hl}[1]{{#1}}

\newcommand{\abs}[1]{\left\lvert{#1}\right\rvert}
\newcommand{\norm}[1]{{\left\lVert{#1}\right\rVert}}

\renewcommand{\b}[1]{\langle#1|} 
\renewcommand{\k}[1]{|#1\rangle} 
\newcommand{\com}[2]{\left[#1,#2\right]}
\newcommand{\proj}[1]{|#1\rangle\langle #1|} 
\newcommand{\rmi}{\mathrm{i}}

\newcommand{\R}{\rangle}
\renewcommand{\L}{\langle}

\def\tr{\mbox{tr}}

\usepackage{url}
\usepackage{tikz}
\usepackage{pgfplots}
\pgfplotsset{compat=1.14}
\usepackage{booktabs}
\usepackage{pgfplotstable}
\usepackage{float}
\usepackage{subcaption}
 
\usepackage{multirow}
\usepackage{bbold}
\usepackage{listings}

\usepackage{enumerate}
\usepackage{latexsym}
\usepackage{amsfonts, amsmath, bbm, amssymb, amsthm}
\usepackage{epic}
\usepackage{braket}
\usepackage{bbm}
\newcommand{\ketbra}[2]{\ket{#1}\bra{#2}}

\begin{document}

\title{Optical nuclear electric resonance as single qubit gate for trapped neutral atoms}

\author{Johannes K. Krondorfer}%
\email{johannes.krondorfer@gmail.com}
\affiliation{ 
Institute of Experimental Physics, Graz University of Technology, Petersgasse 16, 8010 Graz, Austria}

\author{Sebastian Pucher}
\thanks{Current address: Atom Computing, Inc., 2500 55th St, Boulder, Colorado 80301, United States of America}
\affiliation{Max-Planck-Institut für Quantenoptik, Hans-Kopfermann-Straße 1, 85748 Garching, Germany}
\affiliation{%
  Munich Center for Quantum Science and Technology,
  80799 M{\"u}nchen, Germany}

\author{Matthias Diez}
\affiliation{
Institute of Experimental Physics, Graz University of Technology, Petersgasse 16, 8010 Graz, Austria}

\author{Sebastian Blatt}
\affiliation{Max-Planck-Institut für Quantenoptik, Hans-Kopfermann-Straße 1, 85748 Garching, Germany}
\affiliation{%
  Munich Center for Quantum Science and Technology,
  80799 M{\"u}nchen, Germany}
\affiliation{%
  Fakult{\"a}t f{\"u}r Physik,
  Ludwig-Maximilians-Universit{\"a}t M{\"u}nchen,
  80799 M{\"u}nchen, Germany}

\author{Andreas W. Hauser}%
\email{andreas.w.hauser@gmail.com}
\affiliation{ 
Institute of Experimental Physics, Graz University of Technology, Petersgasse 16, 8010 Graz, Austria}

\date{\today}

\begin{abstract}
The precise control of nuclear spin states is crucial for a wide range of quantum technology applications. Here, 
we propose a fast and robust single-qubit gate in $^{87}$Sr, utilizing the concept of optical nuclear electric resonance (ONER). ONER exploits the interaction between the quadrupole moment of a nucleus and the electric field gradient generated by its electronic environment, enabling spin level transitions via amplitude-modulated laser light.
We investigate the hyperfine structure of the 5s$^2$~$^1S_{0}\rightarrow{}$~5s5p~$^3P_1$ optical transition in neutral $^{87}$Sr, and identify the magnetic field strengths and laser parameters necessary to drive spin transitions between the $m_I$ = -9/2 and $m_I$ = -5/2 hyperfine levels in the ground state.
Our simulations show that ONER could enable faster spin operations compared to the state-of-the-art oscillations in this 'atomic qubit'. Moreover, we show that spin-flip operations exceeding 99.9\% fidelity can be performed
even in the presence of typical noise sources.
These results pave the way for significant advances in nuclear spin control, opening new possibilities for quantum memories and other quantum technologies.
\end{abstract}


\maketitle

\section{Introduction}
The precise control of nuclear spin states is essential for a wide range of applications in quantum information processing~\cite{daley_quantum_2008, suter_spins_2008, jones_quantum_2011}, metrology~\cite{ludlow_optical_2015}, and the study of fundamental physical phenomena~\cite{Bloch_ultracold_2008,Safronova_newphysics_2018}. Traditional techniques for nuclear spin manipulation, such as nuclear magnetic resonance (NMR)~\cite{vandersypen_nmr_2005}, rely on radio frequency pulses to induce transitions between nuclear spin states. However, these methods face limitations in terms of spatial resolution, especially in atomic and molecular systems~\cite{ma_universal_2022}. Driven by the need for enhanced spatial quantum control, alternative methods for the manipulation of nuclear spins have been explored, and two-photon Raman transitions have become a popular method for the manipulation of ground-state hyperfine qubits~\cite{barnes_assembly_2022, norcia_midcircuit_2023}. 

Recently, we have proposed another approach for nuclear spin control named `optical nuclear electric resonance' (ONER).\cite{krondorfer_nuclear_2023,krondorfer_optical_2024} It exploits the interaction between the quadrupole moment of a nucleus and the electric field gradient generated by its electronic environment. By periodic modulation of the latter, e.g., via amplitude-modulated laser light, it is possible to induce nuclear spin transitions. ONER enables precise nuclear spin control with high spatial resolution, leveraging optical wavelengths that are much shorter than those used in conventional NMR techniques. Moreover, ONER only requires the application of a single amplitude-modulated laser beam to flip nuclear spins, which can be implemented with an electro-optical modulator.

Within the field of quantum computing, protocols based on nuclear spin manipulation have recently achieved significant advancements~\cite{reichardt_logical_2024}. Nuclear spin qubits have been experimentally demonstrated in neutral atoms, the leading candidates for scalable qubit arrays. Alkaline-earth and alkaline-earth-like atoms are particularly attractive due to their long-lived qubit states, the theoretically well-understood hyperfine structure~\cite{boyd_nuclear_2007}, and the ability to precisely control their interactions with external fields~\cite{saffman_quantum_2016, henriet_quantum_2020}. These properties have led to experimental demonstrations of high-fidelity gates~\cite{noguchi_quantum_2011, jenkins_ytterbium_2022, ma_universal_2022, muniz_high_2024}, erasure conversion~\cite{ma_high_2023}, and midcircuit operations~\cite{lis_midcircuit_2023, huie_repetitive_2023, norcia_midcircuit_2023}. Our target system, the $^{87}$Sr atom, features a nuclear spin which couples only weakly to the environment, making it an excellent candidate for quantum information storage. However, one limitation so far is the achievable Rabi frequencies. Experimentally, Rabi frequencies of just 1.2 kHz have been demonstrated~\cite{barnes_assembly_2022}, which are significantly slower than those achieved with nuclear qubits in e.g. $^{171}$Yb~\cite{jenkins_ytterbium_2022}. Despite this, $^{87}$Sr offers the compelling advantage of coherence times up to several tens of seconds~\cite{barnes_assembly_2022} \--- surpassing the coherence times reported for $^{171}$Yb by far~\cite{jenkins_ytterbium_2022}.
This robustness, combined with the favorable quadrupole moment of the $^{87}$Sr isotope and the presence of a suitable optical transition, makes strontium a prime candidate for future experimental realizations of ONER.
 
In this paper, we show the feasibility of ONER in $^{87}$Sr in a Paschen-Back-like regime, and compute laser parameters and magnetic fields that enable fast and robust spin manipulation in the ground state. We show that fidelities exceeding 99.9\% can be achieved even in the presence of typical noise sources.
Thereby, we show that ONER has the potential to enable faster Rabi oscillations, making it an intriguing alternative to existing methods
, which may translate into faster quantum logics in the long run.

Our article is structured as follows. We start with a general discussion of ONER and continue with a brief overview of the properties of the (5s$^2$)~$^1S_0 \rightarrow$~(5s5p)~$^3P_1$ optical transition in $^{87}$Sr. This is followed by a reformulation of the hyperfine Hamiltonian, a discussion of nuclear quadrupole interaction, and a description of the ONER protocol specifically for $^{87}$Sr. 
We then determine the optimal magnetic field conditions and laser parameters for `qubit' transitions between the $m_I = -9/2$ and $m_I = -5/2$ hyperfine levels of the electronic ground state. After the introduction of our computational simulation technique, we use it to study the coupling between the nuclear quadrupole moment and the electronic environment. The discussion and interpretation of our results, also supported by the application of Floquet theory, is then followed by a stability analysis with respect to experimental parameter settings and laboratory requirements.

\section{The principle of ONER}
\begin{figure}[!t]
    \centering
    \includegraphics[width=0.5\textwidth]{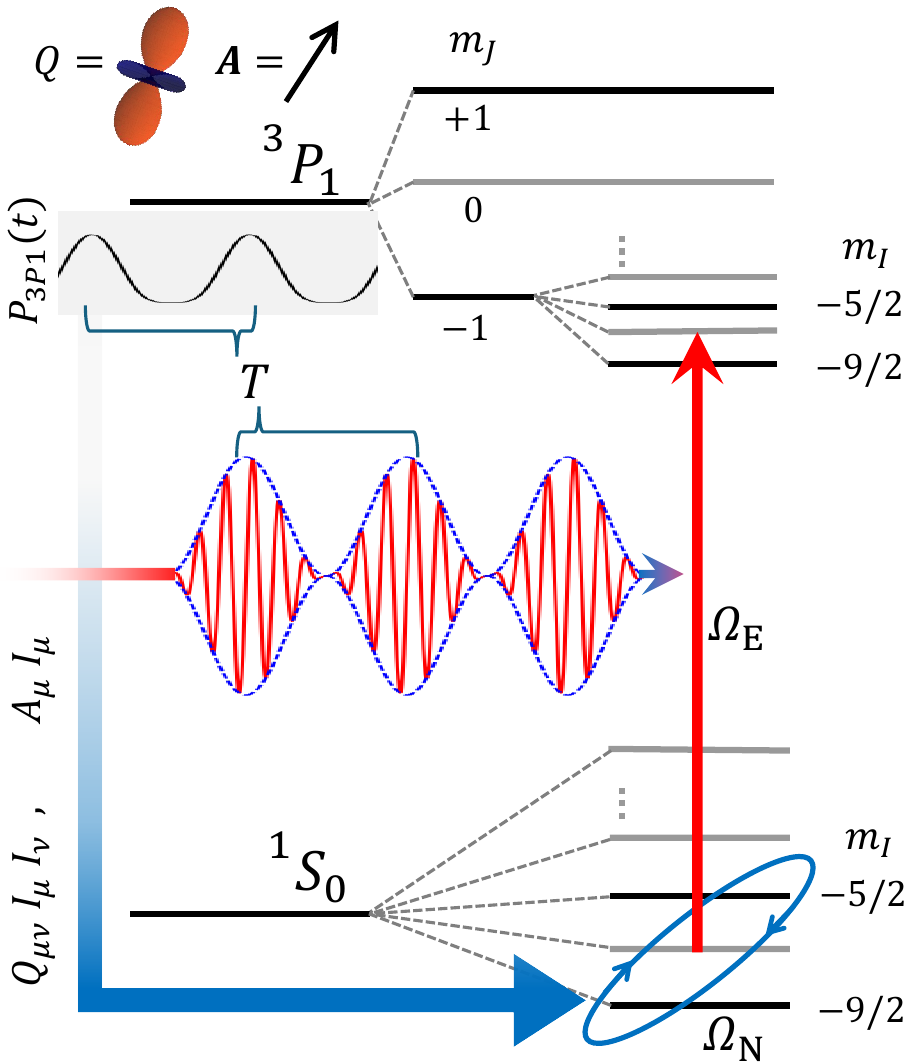}
    \caption{Schematic illustration of the optical nuclear electric resonance (ONER) protocol applied to the level structure of the (5s$^2$)~$^1S_0 \rightarrow$~(5s5p)~$^3P_1$ optical transition in $^{87}$Sr, in a Paschen-Back-like regime. The most relevant levels for ONER are plotted in black, less relevant levels are illustrated in gray. A detuned, amplitude-modulated laser field with period $T$ and electronic Rabi frequency $\Omega_\mathrm{E}/2\pi$ drives the system, resulting in an adiabatically modulated occupation of the electronically excited state, illustrated in the gray box. In the $^1S_0$ ground state, the magnetic nuclear spin quantum number $m_I$ remains well-defined. However, in the $^3P_1$ excited states, the non-zero nuclear quadrupole interaction (NQI) tensor $Q$ (represented by an 'orbital-like' tensor plot surface) and the total magnetic dipole moment $\bm{A}$ of the electrons (represented by a vector) couples to the nuclear spin states. A modulated occupation of the excited state, $P_{3P1}(t)$, leads to a modulation of the NQI tensor and the magnetic dipole moment vector of the electrons, which couple to the nuclear spin $I$ via quadrupole interaction and magnetic dipole interaction, respectively. This enables hyperfine nuclear spin transitions in the $^1S_0$ ground state with a nuclear Rabi frequency $\Omega_\mathrm{N}/2\pi$, illustrated by the blue ellipse enclosing the qubit spin levels.}
    \label{fig:oner schematic}
\end{figure}

We start with a general description of ONER before we discuss its applicability to the $^{87}$Sr system. The basic idea is to exploit a difference between the nuclear quadrupole interaction in different electronic states to drive nuclear spin transitions, achieved by an appropriate time-modulation of electronic state populations. A schematic illustration of the principle, as it may be realized in $^{87}$Sr, the target system of this article, is provided in Figure~\ref{fig:oner schematic}. It shows a part of the hyperfine structure of the (5s$^2$)~$^1S_0 \rightarrow$~(5s5p)~$^3P_1$ optical transition in a magnetic field. The ONER protocol requires the application of an amplitude-modulated laser field to an atomic or molecular system. The laser frequency is chosen to be suitably detuned from the transition energy between the electronic ground state and an excited state. Ideally, the excited state has a strong hyperfine interaction, while the ground state has no hyperfine interaction, thus providing well-defined nuclear spin states. If the amplitude modulation is slow compared to the electron dynamics, the electronic system will follow this modulation adiabatically, i.e. the time dependence of the excited state occupation is inherited from the amplitude modulation of the laser signal (illustrated below the excited state in Figure~\ref{fig:oner schematic}). This periodic modulation of the excited state occupation leads to a periodic modulation of the electric field gradient (EFG), and thus of the nuclear quadrupole interaction (NQI) (illustrated by an `orbital-like' tensor plot surface in the excited state), which can drive nuclear spin transitions by coupling nuclear spin and EFG at the position of the nucleus. Also, the total magnetic dipole moment of the electrons (illustrated by a vector in Figure~\ref{fig:oner schematic}) is modulated by the excited state occupation, leading to additional nuclear spin interactions.
The period $T$ of the amplitude modulation is adjusted such that it allows for specifically selected transitions between hyperfine levels in the ground state. Depending on the structure of the NQI tensor and the total magnetic dipole moment of the electrons, $\Delta m_I = \pm1$ and $\Delta m_I = \pm2$ transitions can be driven. In the following sections, we will focus on the $m_I = -9/2$ to $m_I = -5/2$ transitions in the ground state. A brief summary of the theory behind ONER is provided in Appendix~\ref{app:ONER}.

\section{The $^1S_0$ and $^3P_1$ states in $^{87}\text{Sr}$}\label{sec:87Sr}
In the following sections, we show that the (5s$^2$)~$^1S_0$ $\rightarrow$ (5s5p)~$^3P_1$ transition in $^{87}$Sr is particularly well suited for ONER. We first investigate the dependence of the respective states on an external magnetic field and discuss the important parameters of the system. 

\begin{figure}[!ht]
    \centering
    \includegraphics[width=0.5\textwidth]{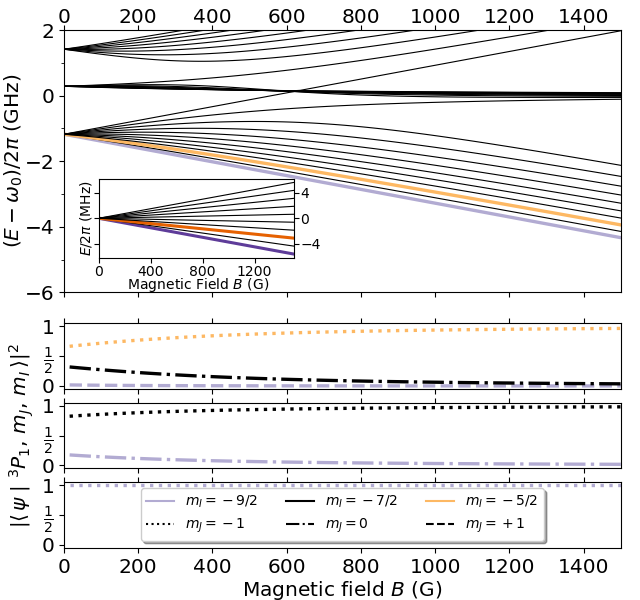}
    \caption{Magnetic field dependence of the hyperfine states of the $^1S_0$ and $^3P_1$ states. The upper plot displays the energy levels in the ground 5s$^2$~$^1S_0$ state (small inset) and the excited 5s5p~$^3P_1$ state of $^{87}$Sr for different magnetic fields $B$. The highlighted states with the lowest and third lowest energy both in the ground state and in the excited state are of main interest in our study. The lower plot displays the projections of the three lowest energy states in the excited $^3P_1$ manifold of $^{87}$Sr (one graph for each state, lowest energy state at the bottom) onto the $\ket{^3P_1,m_J,m_I}$ basis vectors. We only find a small mixing of the quantum numbers $(m_J,m_I)$ in the $^3P_1$ state, indicating that they are good quantum numbers. Only the non-zero components are shown. Line style and line color conventions introduced in this plot are used throughout the manuscript.} \label{fig:energy levels}
\end{figure}

Note that $^{87}$Sr has a nuclear spin quantum number of $9/2$, thus a total of ten nuclear spin states. The $^1S_0$ ground state of $^{87}$Sr has no hyperfine interaction and a total electron angular momentum of zero. Therefore, the magnetic field dependence is just given by the Zeeman interaction of the nuclear spin and the external magnetic field, as shown in the inset of Figure~\ref{fig:energy levels}. For the $^3P_1$ state, the hyperfine interaction of nuclear spin and total angular momentum of the electrons gives rise to a non-trivial Breit-Rabi diagram, as shown in Figure~\ref{fig:energy levels}. It displays a total of 30 states -- 10 spin states multiplied by 3 electronic total angular momentum states. In the further discussion, we will be mostly interested in the lowest and third-lowest lying states of the ground and excited state manifolds highlighted in Figure~\ref{fig:energy levels}.

Considering the subsystem of $^1S_0$ and $^3P_1$ states, the total atomic Hamiltonian $H_\mathrm{A}$ can be expressed as
\begin{equation} \label{eq:total Atomic Hamiltonian}
    H_\mathrm{A} = H_\mathrm{E} + H_\mathrm{Z} + H_\mathrm{HF}\;,
\end{equation}
with the electronic Hamiltonian $H_\mathrm{E}$, the Zeeman Hamiltonian $H_\mathrm{Z}$ and the hyperfine Hamiltonian $H_\mathrm{HF}$. The electronic Hamiltonian represents the electronic excitation (5s$^2$)~$^1S_0$ $\rightarrow$ (5s5p)~$^3P_1$ centered at 689~nm. Its corresponding energy~$\omega_0/2\pi$, expressed in units of GHz, sets the zero point in Figure~\ref{fig:energy levels}. This part of the total Hamiltonian can be written as
\begin{equation}\label{eq:total Sr Hamiltonian electronic}
    H_\mathrm{E} = \omega_0\;\proj{^3P_1} \otimes \mathbb{1}\;,
\end{equation}
with $\hbar{}=1$ for notational convenience.
The first part of the tensor product in Equation~\eqref{eq:total Sr Hamiltonian electronic} covers the electronic, the second covers the nuclear degrees of freedom. The tensor product structure will not be made explicit in the further discussion.
Note that the excited state decays into the ground state with a rate of $\Gamma = 2\pi \times 7.48\;\mathrm{kHz}$,\cite{barnes_assembly_2022,heinz2020gamma} which we include in our simulations as discussed in Section~\ref{sec:full simulation}.

The Zeeman Hamiltonian $H_{\rm Z}$ captures the interaction of the electrons and the nucleus with an external magnetic field $\bm{B} = B_0 \bm{e}_z$ oriented along the $z$-direction. It is given by
\begin{equation}\label{eq:total Sr Hamiltonian zeeman}
    H_\mathrm{Z} = \left(g_J \mu_0 \hat{\bm{J}} - g_I \mu_\mathrm{N} \hat{\bm{I}} \right) \cdot \bm{B}\;,
\end{equation}
with $\hat{\bm{J}}$ denoting the total angular momentum operator of the electrons and $\hat{\bm{I}}$ the spin angular momentum operator. They are multiplied by the Bohr magneton $\mu_0$, the nuclear magneton $\mu_\mathrm{N}$, and by their corresponding g-factors $g_I = -1.0928$ and 
\begin{equation}
g_J = 1 + \frac{J(J+1) + S(S+1) - L(L+1)}{2J(J+1)} = \frac{3}{2}\;,
\end{equation}
respectively.\cite{stone_table_2005}

The last term in Equation~\ref{eq:total Atomic Hamiltonian}, the hyperfine Hamiltonian $H_\mathrm{HF}$, contains the magnetic dipole and electric quadrupole interaction of the electrons and the nucleus. In general, it can be written as
\begin{equation}\label{eq:total Sr Hamiltonian hyperfine}
    H_\mathrm{HF} = A \hat{\bm{I}}\cdot\hat{\bm{J}} + Q\frac{ \frac{3}{2}\hat{\bm{I}}\cdot\hat{\bm{J}} \left( 2\hat{\bm{I}}\cdot\hat{\bm{J}} + 1 \right) - \hat{\bm{I}}^2 \hat{\bm{J}}^2}{2IJ(2I-1)(2J-1)}\;,
\end{equation}
with a hyperfine constant $A=2\pi\times-260$~MHz and a quadrupole constant $Q=2\pi\times-35$~MHz.\cite{zu_putlitz_bestimmung_1963}

For the simulation, we choose a representation in the basis $\ket{n,m_J,m_I}$, with $n\in\{^1S_0,^3P_1\}$, $m_J = 0$ for the $^1S_0$ state and $m_J\in\{\pm 1, 0\}$ for the $^3P_1$ state. The magnetic quantum number $m_I$ of the nuclear spin runs from -9/2 to 9/2.  In general, the eigenstates of the total atomic Hamiltonian $H_\mathrm{A}$ are non-trivial in the chosen basis. In our qubit system, this is clearly the case in the excited state, where the $(m_J,m_I)$ are not good quantum numbers due to the combined Zeeman and hyperfine interactions. Note that this is not a flaw but a necessity of ONER, since we intend to drive well-defined nuclear spin transitions within the ground-state manifold, but use controlled spin mixing in the excited state to drive them. Nevertheless, $(m_J,m_I)$ need to be sufficiently good quantum numbers to avoid undesired spin mixing.
For the three energetically lowest states in the $^3P_1$ manifold, already moderate magnetic field strengths are sufficient to enter a Paschen-Back-like regime with $(m_J,m_I)$ as suitable quantum numbers. This is illustrated in Figure~\ref{fig:energy levels}, where the overlap of the three lowest eigenstates of the $^3P_1$ manifold with the $\ket{^3P_1,m_J,m_I}$ basis is shown. For our purpose, a moderate magnetic field strength of $200$ Gauss is already sufficient to enable the ONER protocol. Asymptotically, for large $B$, the three lowest eigenstates of the $^3P_1$ manifold are clearly defined as -9/2, -7/2, and -5/2 states with respect to $m_I$, and as -1 states with respect to $m_J$. Conventions regarding line style and color, as introduced in Figure~\ref{fig:energy levels}, are used throughout the manuscript.

\section{Optical nuclear electric resonance}
\subsection{Hyperfine Hamiltonian and nuclear quadrupole interaction}
In the context of ONER, it is convenient to phrase the hyperfine interaction, given in Equation~\eqref{eq:total Sr Hamiltonian hyperfine}, differently. First, we note that two contributions appear: a linear and a quadratic interaction in the spin and angular momentum operators.

The linear term arises from the interaction of $\hat{\bm{\mu}}_I$, the magnetic moment of the nucleus, with the magnetic field $\hat{\bm{B}}_\text{el}$ generated by the surrounding electrons, yielding $H_{\mathrm{HF},\text{lin}}=\hat{\bm{\mu}}_I\cdot \hat{\bm{B}}_\text{el}$. Both are vector operators in their respective Hilbert spaces, and can therefore be related by the Wigner-Eckart theorem~\cite{wigner1959group} to the nuclear spin $\hat{\bm{I}} \sim \hat{\bm{\mu}}_I$ and the total angular momentum of the electrons $\hat{\bm{J}} \sim \hat{\bm{B}}_\text{el}$, respectively. Combining the proportionality constants into a single linear hyperfine constant $A$ yields the linear part of the hyperfine Hamiltonian, written as $\hat{\bm{A}}\cdot\hat{\bm{I}}$, with the total magnetic dipole moment of the electrons $\hat{\bm{A}}:= A\hat{\bm{J}}$.

Likewise, we can rephrase the quadratic interaction of the hyperfine Hamiltonian, the key feature of any nuclear electric resonance protocol, including ONER. This term is caused by the interaction of the electric quadrupole moment of the nucleus $\hat{\mathcal{Q}}_{mn}$ and the electric field gradient (EFG) at the position of the nucleus $\hat{\Phi}_{mn}$. Both are symmetric traceless second-rank tensors in their respective Hilbert spaces. The corresponding Hamiltonian is given by
\begin{equation}\label{eq:HF quad}
    H_{\text{HF,quad}} = \frac{1}{6} \tr\{\hat{\mathcal{Q}} \hat{\Phi}\} =  \frac{1}{6} \sum_{\mu\nu}\hat{\mathcal{Q}}_{\mu\nu} \hat{\Phi}_{\mu\nu}\;.
\end{equation}
This Hamiltonian can be derived from a second-order Taylor expansion of the atomic Hamiltonian for a finite-sized nucleus.\cite{krondorfer_nuclear_2023}.

By employing the Wigner-Eckart theorem, the nuclear quadrupole tensor and the EFG tensor can be related to the nuclear spin operator and total electronic angular momentum operator, respectively. This gives
\begin{align}
\begin{split}
    \hat{\mathcal{Q}}_{nm} &= \frac{q}{I(2I-1)} \left( \frac{3}{2} (\hat{I}_n \hat{I}_m + \hat{I}_m \hat{I}_n) - \delta_{nm} \hat{\bm{I}}^2  \right)\;, \\
    \hat{\Phi}_{nm} &= \frac{\phi}{J(2J-1)} \left( \frac{3}{2} (\hat{J}_n \hat{J}_m + \hat{J}_m \hat{J}_n) - \delta_{nm} \hat{\bm{J}}^2  \right)\;,
\end{split}
\end{align}
with the scalar nuclear quadrupole moment $q$ and the scalar electric field gradient $\phi$. Note that only nuclei with nuclear spin $I > \frac{1}{2}$ can have a non-vanishing quadrupole moment and only electronic states with $J > \frac{1}{2}$ can have non-vanishing EFG, as indicated by the proportionality constants. 

Inserting these representations in Equation~\eqref{eq:HF quad} and reordering, we obtain the typical form of the hyperfine interaction Hamiltonian in the context of atomic physics by setting $Q = q\phi$. Another representation, typically used in nuclear magnetic resonance spectroscopy, is obtained by just replacing $\hat{\mathcal{Q}}$ in Equation~\eqref{eq:HF quad} and exploiting the symmetry properties of the electric field gradient tensor. This yields
\begin{equation} \label{eq:NSI}
    H_{\text{HF,quad}} = \sum_{\mu\nu} \hat{I}_\mu \hat{Q}_{\mu\nu} \hat{I}_\nu\;,
\end{equation}
with $\hat{Q}_{\mu\nu} = \frac{q}{2I(2I-1)} \hat{\Phi}_{\mu\nu}$ as the nuclear quadrupole interaction (NQI) tensor. Note that the NQI tensor is an operator in the Hilbert space of the electronic degrees of freedom, and does not depend on the nuclear spin degrees of freedom. Thus, the NQI tensor is fully determined by the electronic configuration, or more specifically, by the total electronic angular momentum operator $\hat{\bm{J}}$ and its time-dependent expectation value. The same holds for the magnetic dipole moment $\hat{\bm{A}}$ of the electrons.

\subsection{Quadrupole interaction for nuclear spins}\label{sec:quadrupole interaction}
To identify suitable parameters for nuclear spin transitions via ONER, we investigate the nuclear quadrupole interaction Hamiltonian of a quadrupolar nucleus placed in an external magnetic field $\bm{B}=B_0 \bm{e}_z$ in $z$-direction, within a known adiabatically controlled electronic environment, i.e. a well-defined electric field gradient (EFG). We indicate the known EFG by replacing $\hat{\Phi}$ with its time-dependent expectation value $\braket{\hat{\Phi}}$, leading, according to Equation~\eqref{eq:NSI}, to an averaged NQI tensor~$\braket{\hat{Q}}_{\mu\nu}$. The corresponding total Hamiltonian, comprising Zeeman and quadrupole interactions, can be written as
\begin{align} \label{eq:HbHq}
H = H_B + H_Q = -\gamma_\mathrm{n} B_0 \hat{I}_z + \sum_{\mu\nu} \hat{I}_\mu \braket{\hat{Q}}_{\mu\nu} \hat{I}_\nu\;,
\end{align}
with $\gamma_\mathrm{n}$ denoting the gyromagnetic moment of the nucleus and the NQI tensor as given above.

First, if the NQI is sufficiently small compared to the Zeeman interaction, then the time-independent part of the NQI will lead to additional line shifts of the Zeeman energies and therefore lift the degeneracy of the transition energies. In good approximation, the eigenstates of the total Hamiltonian remain well-defined in the eigenbasis of the Zeeman-Hamiltonian comprising the orientational nuclear spin states $\k{m_I}$ for a fixed spin quantum number $I$. The resulting energy correction is given in Appendix~\ref{app:quadrupole energy}. Note that the correction of the transition energies opens the possibility to address specific transitions individually, which is not possible in the case of equidistant Zeeman-splitting alone. For the $^1S_0 \rightarrow$$^3P_1$ transition discussed here, individual control is enhanced by the hyperfine interaction in the excited state, as we discuss in Section~\ref{sec:full simulation}.

Second, a time-dependent variation of the NQI tensor allows for transitions between nuclear spin states. Since the NQI is quadratic in the spin operator $I$, it allows for driving transitions with $\Delta m_I = \pm 1, \pm 2$. The corresponding transition amplitudes are given as~\cite{krondorfer_nuclear_2023}
\begin{align} \label{eq:trans1}
\begin{split}
g_{m_I \rightarrow m_I-1}(t) &= \alpha_{m_I-1 \leftrightarrow m_I} \left( \braket{\hat{Q}}_{xz}(t) + \rmi \braket{\hat{Q}}_{yz}(t) \right)\;,\\
\alpha_{m_I-1 \leftrightarrow m_I} &= \frac{1}{2} \abs{2 m_I -1} \sqrt{I(I+1) - m_I (m_I - 1)}\;,
\end{split}
\end{align}
\begin{align} \label{eq:trans2}
\begin{split}
g_{m_I \rightarrow m_I-2}(t) &= \beta_{m_I-2 \leftrightarrow m_I} \times \\&\left( \braket{\hat{Q}}_{xx}(t) - \braket{\hat{Q}}_{yy}(t) +  2\rmi \braket{\hat{Q}}_{yx}(t) \right)\;, \\
\beta_{m_I-2 \leftrightarrow m_I} &= \frac{1}{4} \sqrt{ \left( I(I+1) - (m_I-1) ( m_I -2) \right) } \\ &\qquad\quad \times \sqrt{ \left( I(I+1) - m_I ( m_I -1) \right)}\;.
\end{split}
\end{align}
The transition amplitudes for the reverse transition are given by the complex conjugate of the above expressions. They reveal that the structure of the quadrupole interaction tensor determines the type and strength of the possible transitions, and is therefore a crucial element of the ONER protocol.

\subsection{Electric field gradients for atomic states}\label{sec:EFG for atoms}
For atomic systems, the structure of~$\braket{\hat{Q}}_{\mu\nu}$, the averaged NQI tensor, can be deduced from the respective electronic state, i.e. the $(J,m_J)$ quantum numbers. Note, however, that pure states of the form $\ket{\psi} = \ket{J,m_J}$ give rise to a cylindrical NQI tensor $\braket{\hat{Q}}=\text{diag}\left(a,a,-2a\right)$, for some constant $a$. This does not allow for transitions between spin states, which is immediate from Equations~\eqref{eq:trans1} and~\eqref{eq:trans2}. Therefore, slight mixtures of states are necessary to enable non-cylindrical NQI tensors. 

For $J < 1$ the electric field gradient vanishes and no NQI occurs. Therefore, $S$-states are a natural choice for the ground state manifold. Other states may be chosen as well, but it is essential to reach the Paschen-Back regime in experimental setups, where the Zeeman interaction dominates over the quadrupole interaction. This ensures that the Zeeman states serve as sufficiently good approximations of the ground state spins. 

A reasonable choice for the electronically excited state is therefore a $P$-state with $J=1$.
For $J=1$, we obtain the possibility of $\Delta m_I = \pm 2$ transitions for state mixtures of $m_J = +1$ and $m_J = -1$, while suppressing $\Delta m_I = \pm1$ transitions. For mixtures of $m_J = \pm 1$ and $m_J = 0$, also $\Delta m_I = \pm1$ transitions are enabled. However, $\Delta m =\pm1$ transitions are also affected by the linear hyperfine interaction. If this interaction is too large with respect to the quadrupole interaction, then these transitions are compromised and of reduced stability. 

Therefore, $\Delta m_I = \pm2$ transitions are an ideal candidate to isolate nuclear quadrupole interactions and thereby driven transitions while suppressing linear hyperfine interaction. The mixture of states is provided by the quadrupole interaction itself and second-order perturbation in linear hyperfine interaction, and through the application of a laser field polarized perpendicular to the quantization axis. Note that the application of circularly polarized laser light will not lead to transitions, as it couples only to a single excited state, leading to a cylindrical NQI tensor.

\section{Results for ONER in $^{87}\text{Sr}$}\label{sec:full simulation}
Having introduced ONER in general, we can now apply the protocol to the (5s$^2$)~$^1S_0$ $\rightarrow$ (5s5p)~$^3P_1$ transition in $^{87}$Sr.
We begin with a discussion of laser parameters suitable for transitions between the $m_I=-9/2$ and $m_I=-5/2$ hyperfine ground state. The reason for choosing the $m_I=-5/2$ state instead of the $m_I=-7/2$ state is twofold: First, it is a $\Delta{}m_I=2$ transition, with all the properties mentioned above.
Second, it is less affected by the large hyperfine interaction in the excited state manifold, which is not the case for the $m_I = -9/2$ to $m_I = -7/2$ transition. The involvement of hyperfine interaction would make the procedure unstable and introduce an undesired, additional coupling of the spin states in the excited state manifold.

We start with an overall discussion of the simulation setup in Section~\ref{sec:set up}, defining the total Hamiltonian and suitable laser parameters, and provide simulation results and interpretations in Section~\ref{sec:results}.

\subsection{Simulation setup and total Hamiltonian} \label{sec:set up}
To drive transitions between the $m_I=-9/2$ and $m_I=-5/2$ ground state, we place the atom in an amplitude-modulated, linearly polarized laser field of polarization angle~$\theta=\pi/2$ with respect to the quantization axis $z$, frequency~$\omega$, and a time-dependent electronic Rabi frequency
\begin{equation}
    \Omega_\mathrm{E}(t)/2\pi = \frac{\Omega_\mathrm{E}/2\pi}{2} \left(1 - \cos\left(\frac{2\pi t}{T}\right)\right)\;,
\end{equation}
with an amplitude modulation period $T$. Note that the intensity of the excitation laser is proportional to the square of the Rabi frequency of the electronic transition. In our simulations, we use the electronic Rabi frequency as a parameter representing the experimentally tunable laser intensity. Details on the conversion of electronic Rabi frequency to laser intensity are provided in Appendix~\ref{app: dipole moment}.
The total Hamiltonian is then given by the sum of the atomic Hamiltonian $H_\mathrm{A}$ and the atom laser field interaction Hamiltonian $H_\mathrm{AF}$. We transform the system into the rotating frame and apply the rotating wave approximation (RWA). This results in the Hamiltonian
\begin{equation}\label{eq:total time dependent Hamiltonian}
    H = H_\mathrm{E}' + H_\mathrm{Z} + H_\mathrm{HF} + H_\mathrm{AF}\;,
\end{equation}
with
\begin{equation}
    H_\mathrm{E}' = -\Delta\;\proj{^3P_1}\otimes\mathbb{1}\;,
\end{equation}
where $\Delta := \omega_0 - \omega$ is the detuning of the laser frequency from the central transition frequency $\omega_0$. The atom field interaction Hamiltonian, determined by the electronic Rabi frequency~$\Omega_\mathrm{E}(t)/2\pi$ and the polarization angle~$\theta$, can be written as
\begin{align}
\begin{split}
    H_\mathrm{AF}(t) = \frac{1}{2} \Omega_\mathrm{E}(t)\; ( 
    \cos(\theta) \ketbra{^1S_0}{^3P_1,0} &\otimes \mathbb{1}\,+ \\
    \frac{\sin(\theta)}{\sqrt{2}} \ketbra{^1S_0}{^3P_1,-1} &\otimes \mathbb{1}\,- \\
    \frac{\sin(\theta)}{\sqrt{2}} \ketbra{^1S_0}{^3P_1,+1} &\otimes \mathbb{1} + c.c )\;.
\end{split}
\end{align}

We perform simulations with the total time-dependent Hamiltonian given in Equation~\eqref{eq:total time dependent Hamiltonian} to identify parameters that lead to a spin flip between the $m_I = -9/2$ and $m_I = -5/2$ ground states. For this purpose, we choose a laser frequency $\omega$ exactly between the energies of the $\ket{^1S_0,0,-9/2}\rightarrow\ket{^3P_1,-1,-9/2}$ and $\ket{^1S_0,0,-5/2}\rightarrow\ket{^3P_1,-1,-5/2}$ transitions, i.e.
\begin{align}\label{eq: set laser frequency}
\begin{split}
    \omega = \frac{1}{2} ( &( E(^3P_1,-1,-9/2) - E(^1S_0,0,-9/2) ) \\ 
    + &(E(^3P_1,-1,-5/2) - E(^1S_0,0,-5/2)) )\;,
\end{split}
\end{align}
as also illustrated in Figure~\ref{fig:oner schematic}.
This configuration enhances the selectivity of nuclear spin control by suppressing the undesired $m_I=-5/2\leftrightarrow m_I=-1/2$ ground state transition. The suppression arises from multiple factors: the laser field is detuned from the resonance condition for this transition, the effective hyperfine interaction induces an additional perturbative shift of the energy levels, and thus, the frequency of the amplitude modulation is from its optimal value for this transition.
Moreover, choosing the laser frequency in such a way results in an equal detuning for both $m_I=-9/2$ and $m_I=-5/2$ states. This is essential, since both states should experience equal quadrupole coupling.

We perform a series of scans over the period $T$ of the amplitude modulation between $1/100\;\mathrm{\mu s}$ and $5\;\mathrm{\mu s}$ with a resolution of $5\;\mathrm{ns}$, for different magnetic fields $B$ and electronic Rabi frequencies $\Omega_\mathrm{E} / 2\pi$. The results are discussed in the next section.

The bounds for the period of the amplitude modulation are selected such that they comprise the adiabatic regime, i.e. the amplitude modulation is slow enough for the RWA to remain valid, but fast enough to yield large nuclear Rabi frequencies $\Omega_\mathrm{N} / 2\pi$ for the spin system. In other words, several amplitude modulation cycles occur within a single Rabi oscillation of the spin system. We simulate the time evolution of the system for 50~$\mathrm{\mu s}$ using the python module QuTiP~\cite{qutip1,qutip2}, allowing us to resolve nuclear Rabi frequencies of $10\;\mathrm{kHz}$ or higher. As initial state, we choose the $m_I = -9/2$ ground state, and we perform the simulation of the time evolution using a Born-Markov Master equation to account for the decay~$\Gamma$ of the excited state into the ground state manifold. Further details regarding the numerical simulation can be found in Appendix~\ref{app:simulation}.

\hl{Note that with this choice of parameters, the decay of the ${}^3P_1$ manifold is mostly negligible due to its comparatively long lifetime 
($\Gamma^{-1}\!\approx\!20~\mu$s) and the small occupation of the excited state manifold $P_{3P1} \ll 1$. The small admixture of the excited state causes a weak effective decay of the nuclear coherence, scaling approximately as 
$\Gamma_\mathrm{eff}\!\approx\!(\Omega_{\mathrm E}/\delta)^2\Gamma$,\cite{saffman_quantum_2016, ludlow_optical_2015}
analogous to Raman-type transitions, where $\delta\approx \abs{E(^3P_1,-1,-9/2) - E(^3P_1,-1,-5/2)}/2$ is the detuning from the transition, which lies between the two relevant levels in the excited state manifold. Typically, this decay can be neglected if the detuning is sufficiently larger than the intensity. This ensures that the driven dynamics are essentially coherent within the operation time.
}

\subsection{Simulation results and interpretation}\label{sec:results}
\begin{figure*}[!ht]
    \centering
    \begin{subfigure}[b]{0.49\textwidth}
        \centering
        \includegraphics[width=\textwidth]{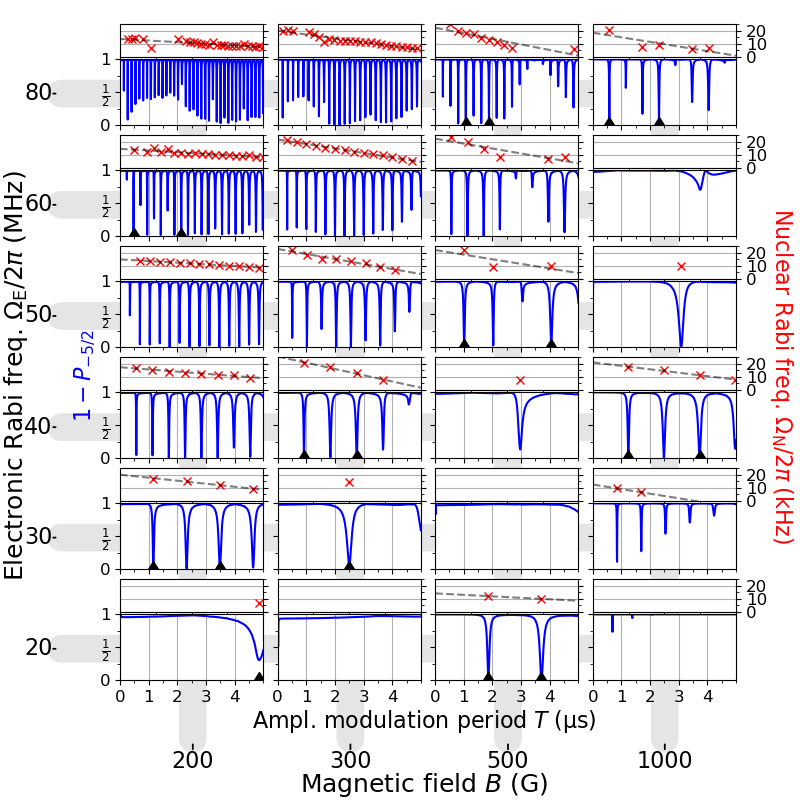}
        \put(-252,240){\small (a)}
        \phantomcaption\label{fig:trans}
    \end{subfigure}
    \begin{subfigure}[b]{0.49\textwidth}
        \centering
        \includegraphics[width=\textwidth]{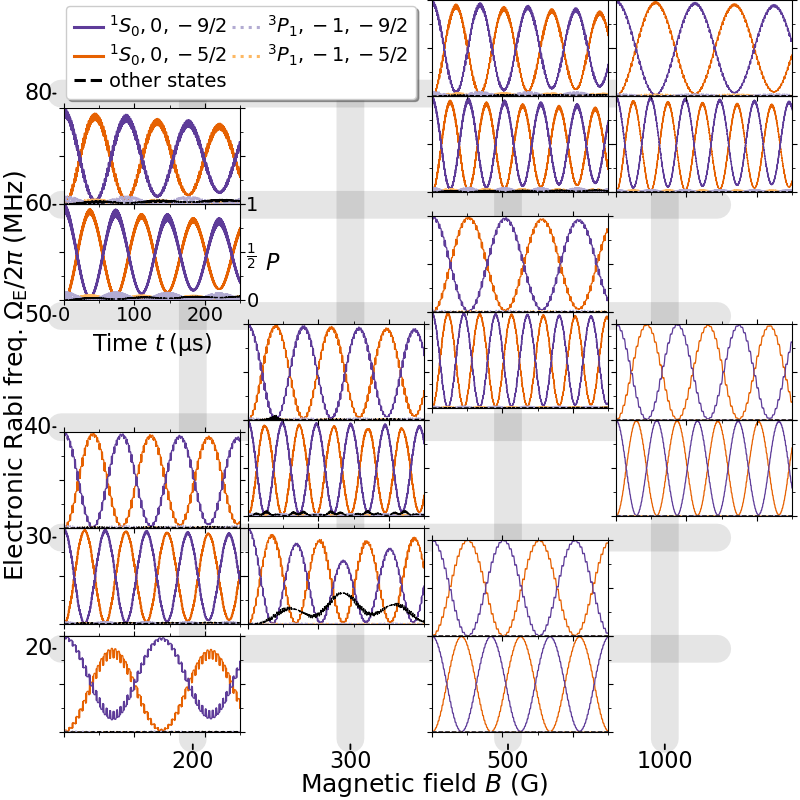}
        \put(-252,240){\small (b)}
        \phantomcaption\label{fig:rabi}
    \end{subfigure}
    \caption{Analysis of Rabi oscillations on the $\ket{^1S_0,0,-9/2}$ to $\ket{^1S_0,0,-5/2}$ transition induced by the ONER method.\newline{}
    (a) Probability $P_{-5/2}$ (blue lines) for flipping the nuclear spin of $^{87}$Sr in the electronic ground state from $m_I=-9/2$ to $m_I=-5/2$, and observed spin Rabi frequencies $\Omega_\mathrm{N}$ for this transition (red crosses), plotted as a function of the modulation period $T$ of the excitation laser. A total of $6\times{}4=24$ graphs are compiled into a single figure, with the outermost tabular arrangement indicating the chosen values of the electronic Rabi frequency $\Omega_\mathrm{E}/2\pi$, and the magnetic field $B$, respectively. The graphs reveal the amplitude modulation periods $T$ at which spin flips occur, identifiable by equidistant dips and characterized by linearly decreasing nuclear Rabi frequencies. Note the similarities in the graphs along diagonals, indicating similar behavior of the excited state occupation under these conditions. Increasing laser intensity, i.e. larger $\Omega_\mathrm{E}/2\pi$, yields sharper and more tightly packed absorption features, yielding less robust transitions with respect to the modulation period $T$. Incomplete transitions, i.e. dips that do not reach zero, are mainly caused by limited resolution in $T$, limited overall simulation time $t\leq50\;\mathrm{\mu s}$, or, in rare cases, by spin mixture. \newline{} (b) Time evolution of the spin level occupation $P$ for selected transitions between the $m_I=-9/2$ and $m_I=-5/2$ levels (absorption features marked by black triangles in \ref{fig:trans}). If two absorption features are selected, the lower modulation period is shown below. The sum over the remaining occupations $m_I\notin\{-9/2,-5/2\}$ (other states) reveals the spin mixing, which occurs if the nuclear Rabi frequency is too large with respect to the effective spin level splitting. We observe high-fidelity Rabi oscillations across various magnetic fields and laser parameters, particularly also for parameters readily achievable in standard laboratories. }\label{fig:trans and rabi} 
\end{figure*}

Now we discuss the results of our simulations, which reveal an intriguing structure of the transition probabilities. Offering amplitude modulation periods $T$ between 0 and $5~\mu{}$s, full flips of the nuclear spin typically appear for several, equidistantly placed values of $T$. Their actual position depends on the magnetic field $B$ and the electronic Rabi frequency of the $^1S_0\rightarrow$$^3P_1$ transition $\Omega_\mathrm{E}/2\pi$. This periodicity is caused by multi-photon transitions, where the amplitude modulation period matches an effective energy difference between the target spin states
\begin{equation}
    \Delta E_{-9/2,-5/2}^\text{eff} = n\frac{2\pi}{T}\,,
\end{equation}
for some $n\in\mathbb{N}$. Due to the time-dependent light shifts generated by the amplitude-modulated laser field, an analytical computation of the effective energy difference is unfeasible. In this section, we focus on the discussion of the numerical results obtained in the simulation of the total system. In Section~\ref{sec:floquet}, we can analyze the behavior with the help of Floquet theory to obtain an effective Hamiltonian for the target subspace and provide an additional explanation for the dynamics of the system. Figure~\ref{fig:trans and rabi} summarizes the numerical results of $6\times{}4$~scans over modulation period~$T$ for $B\in\{200\;\mathrm{G},\,300\;\mathrm{G},\,500\;\mathrm{G},\,1000\;\mathrm{G}\}$ and $\Omega_\mathrm{E}/2\pi\in\{20\;\mathrm{MHz},\,30\;\mathrm{MHz},\,40\;\mathrm{MHz},\,50\;\mathrm{MHz},\,60\;\mathrm{MHz},\,80\;\mathrm{MHz}\}$.

In the left panel, Figure~\ref{fig:trans}, we plot nuclear spin-flip probabilities (blue lines),  $P_{-5/2} = \max_{t\in\tau} \vert\braket{^1S_0,0,-5/2 \vert \psi(t)}\vert^2$, together with the resulting nuclear Rabi frequencies $\Omega_\mathrm{N}/2\pi$ (red crosses) at the absorption features, as a function of the modulation period~$T$. The probabilities are evaluated within a simulation interval $\tau = [0, 50\;\mathrm{\mu s}]$, where $\ket{\psi(t)}$ is the time-dependent wave function of the driven system.  Essentially, the flip probability $P_{-5/2}$ can be interpreted as the $\pi$-gate fidelity at the first flip. The observed nuclear Rabi frequencies show an approximately linear drop from the lowest transition peak to the highest. This is reasonable, as each oscillation of the amplitude modulation 'drives' the system further towards the transition, yielding faster transitions for higher amplitude modulation frequencies. In terms of multi-photon transitions, this behavior is explained by the lower probability of transitions involving more photons.

The peak structure depends on the occupation of the excited state, which is determined by the laser detuning and intensity. In our case, the detuning is fixed for a given magnetic field (see Equation\eqref{eq: set laser frequency}) and increases with increasing magnetic field. Therefore, similar peak structures can be observed along the diagonals. If the laser intensity is too low, such that the occupation of the exited state is too small and therefore also the effective quadrupole interaction, then no transition can be observed on the given timescale.

The depth of the absorption features in Figure~\ref{fig:trans} may differ from~1 mainly due to three reasons. First, and most important for this study, due to spin mixing and excitation into other spin levels of the electronic ground state, which indicates that the hyperfine interaction is still dominant or that additional transitions are driven by the applied laser field, such as the $m_I=-5/2$ to $m_I=-1/2$ transition. Second, due to the limited resolution of 5~ns for the period $T$ of the amplitude modulation in the scan. This limitation appears at large $\Omega_\mathrm{E}/2\pi$ with many, tightly packed and very sharp absorption features, and is easily identified by the somewhat arbitrarily shaped envelopes of the peak series. Third, in cases where the nuclear Rabi frequency~$\Omega_\mathrm{N}/2\pi$ lies below 10~kHz and can no longer be fully resolved within the simulation time window $[0,50\;\mathrm{\mu s}]$. This artifact appears for low $\Omega_\mathrm{E}/2\pi$ or large periods of the amplitude modulation, but is irrelevant for this study, since we are interested in fast spin transitions with $\Omega_\mathrm{N}/2\pi$ larger than 10~kHz. In all of the above cases, the transition peak is of less interest for further investigation, as it will not result in a robust and fast control of the nuclear spins.

The right panel, Figure~\ref{fig:rabi}, shows the simulated nuclear Rabi oscillations for a few selected absorption features marked by black triangles in Figure~\ref{fig:trans}. These plots, featuring a similar grid-like arrangement as their corresponding spin-flip probabilities in terms of parameters $B$ and $\Omega_\mathrm{E}$, represent time evolutions of spin level occupations $P$ (between 0 and 1) over a duration of $250~\mu{}$s. If two absorption features are selected, the lower modulation period is always shown at the bottom. We observe high-fidelity Rabi oscillations across various magnetic field and laser parameters, particularly also for parameters readily achievable in standard laboratories. Good Rabi oscillations are obtained for small amplitude modulations and low occupations of the electronically excited state, i.e., lower intensity or higher magnetic field.

For high intensities and low magnetic fields, the excited state occupation is large, $\approx 0.2$, which entails a fast beat between the two electronic states. As a consequence, a blurred-out line in the time evolution of the system appears due to the limited resolution in the graphics.  Additionally, the finite lifetime of the more populated excited state leads to a pronounced damping of the driven spin oscillations, which is not the case for a low population of the excited state, where coherence is preserved. \hl{This damping, visible in Figure~\ref{fig:rabi}, thus reflects the weak admixture of the short-lived ${}^3P_1$ state. As expected from the scaling $\Gamma_\mathrm{eff}\!\approx\!(\Omega_{\mathrm E}/\delta)^2\Gamma$, the effect becomes noticeable only when the excited-state population exceeds a few percent at strong driving. In the experimentally relevant regime of lower intensities, this induced decay remains negligible, and the oscillation frequency and amplitude are essentially unaffected.
}

\begin{figure}[!ht]
    \centering
    \includegraphics[width=0.5\textwidth]{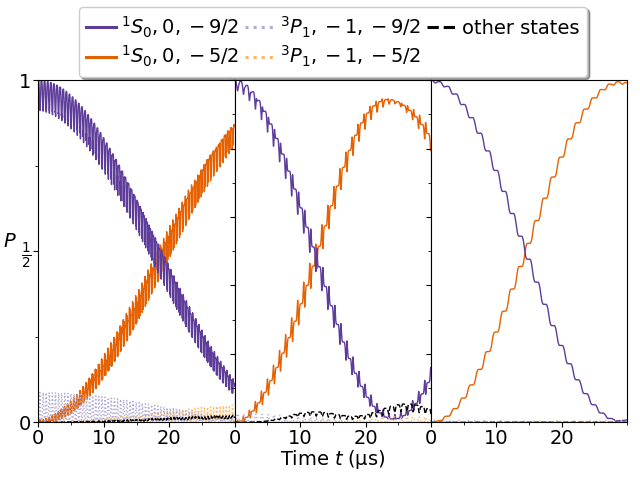}
    \caption{Close-up of the Rabi oscillations for parameter combinations: (200 G, 60 MHz, first peak), (300 G, 40 MHz, first peak), and (1000 G, 40 MHz, first peak) from left to right. The population $P$ is shown over a time interval from 0 to 30~$\mathrm{\mu s}$. At high laser intensity, the excited state is strongly populated. Besides the global Rabi oscillation caused by the amplitude modulation that 'drives' the system further towards the transition, small local oscillations appear. These are caused by the amplitude modulation and the occupation of the excited state, as visible in the left plot. Additionally, hyperfine interaction can lead to oscillations within one external modulation period, due to small mixing of spin states, as can be seen in the middle plot. For a low population of the excited state and a small amplitude modulation period, smooth Rabi oscillations are achieved, as exemplified in the right plot.} \label{fig:rabi close up}
\end{figure}

Moreover, we see that spin mixing can occur if the nuclear Rabi frequency $\Omega_\mathrm{N}/2\pi$ is too large for the effective splitting of the transition energies of the spin system; this happens in the investigated case for a magnetic field of 300 Gauss and the first peak at 30 and 40 MHz electronic Rabi frequency $\Omega_\mathrm{E}/2\pi$. In Figure~\ref{fig:rabi close up}, we provide enlarged, more detailed plots of the time evolution for three selected parameter combinations at higher resolution. From left to right, we present the Rabi oscillations for the parameter settings (200 G, 60 MHz, first peak), (300 G, 40 MHz, first peak), and (1000 G, 40 MHz, first peak).

The more detailed time evolution shows not only the Rabi frequency of the nuclear spin system, but also reveals a fast modulation of very small amplitude. The latter stems from the amplitude modulation with period $T$ and the corresponding occupation of the excited states, as illustrated in the left plot of Figure~\ref{fig:rabi close up}. Furthermore, smaller oscillations can be spotted between periods, particularly at $B=300\;\mathrm{G}$ and $\Omega_\mathrm{E}/2\pi = 40\;\mathrm{MHz}$, which are caused by hyperfine interaction and the corresponding mixture of spin states (see the middle plot of Figure~\ref{fig:rabi close up}). Thus, the time evolution of the total system consists of a slow global time evolution in combination with a fast, local process. The local (stroboscopic) time evolution is approximately constant for the considered states, and its magnitude is much smaller than the amplitude of the global evolution. This is since $m_I$ remains a sufficiently good quantum number, also in the excited state, since the system is in a Paschen-Back-like regime. In case of a weakly occupied electronically excited state, the stroboscopic evolution is constant and small plateaus appear between periods, leading to smooth spin Rabi oscillations (see the right plot in Figure~\ref{fig:rabi close up}), achieved for various laser and magnetic field parameters.

\subsection{Floquet analysis}\label{sec:floquet}
We focus on two selected examples: the third peak of the $B=300\;\mathrm{G}$, $\Omega_\mathrm{E}/2\pi = 40\;\mathrm{MHz}$ transition and the first peak of the $B=500\;\mathrm{G}$, $\Omega_\mathrm{E}/2\pi = 20\;\mathrm{MHz}$ transition. Due to the failure of the Born approximation for singlet/triplet systems (see Appendix~\ref{app:ONER}), a simplified effective analysis is not applicable, although it provides a useful heuristic. However, since the total Hamiltonian of the system in RWA is periodic in time, with period $T$ of the amplitude modulation, Floquet theory can be applied. Details on the principles of Floquet theory can be found in Appendix~\ref{app:Floquet}. In this section, we apply Floquet theory to analyze the structure of the transitions with respect to the amplitude modulation period $T$. 

\begin{figure}[ht]
    \centering
    \includegraphics[width=0.4\textwidth]{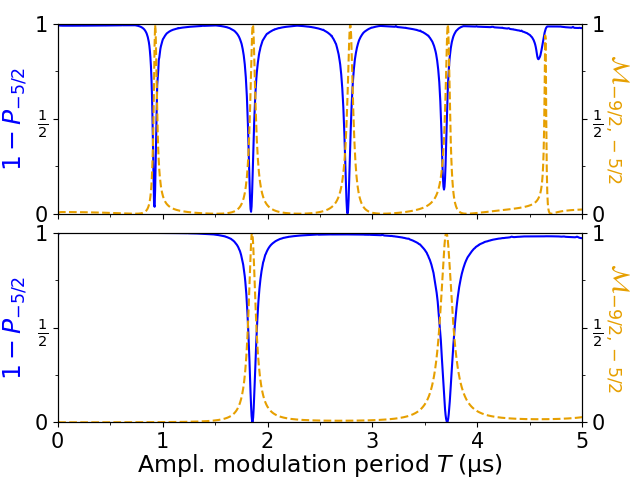}%
    \caption{Comparison of the results from the Floquet analysis with the full time evolution (solid blue line, left axis). The state mixture of $\ket{^1S_0,0,-9/2}$ and $\ket{^1S_0,0,-5/2}$ in the Floquet mode basis is given by Equation~\eqref{eq:state mixture} (dashed orange line, right axis). Two cases are considered: $B=300\;\mathrm{G}$ and $\Omega_\mathrm{E}/2\pi = 40\;\mathrm{MHz}$ (top) and $B=500\;\mathrm{G}$ and $\Omega_\mathrm{E}/2\pi = 20\;\mathrm{MHz}$ (bottom). The peak shape and position are well described by the state mixture measure $\mathcal{M}$ (Equation~\eqref{eq:state mixture}) for the Floquet mode basis.}
    \label{fig:Floquet overlap}
\end{figure}

We calculate the one-period time evolution operator $U(T) = \mathcal{T}\exp\left( -\rmi\int_0^T H(t')\;\mathrm{d}t' \right)$, with $\mathcal{T}$ denoting the time-ordering operator. Since the total Hamiltonian does not commute for different times, we perform a numerical calculation of $U(T)$. The eigenstates of $U(T)$ are the Floquet modes $\ket{n}$. Knowing these Floquet modes and their overlap with the initial wavefunction determines the time evolution for integer multiples of one period, since $U(T)^k=U(kT)$. These Floquet modes reveal the same structure as the full time evolution analyzed in Section~\ref{sec:results} and Figure~\ref{fig:trans}. A complete spin flip occurs only when two Floquet modes form a 50\% mixture of the $m_I=-9/2$ and $m_I=-5/2$ hyperfine ground states, as is illustrated in Figure~\ref{fig:Floquet overlap}.

Figure~\ref{fig:Floquet overlap} compares the dependence of the transitions on $T$ obtained by the full simulation of the system with the predictions based on the overlap of the Floquet modes with the target spin states. The individual absorption features are well resolved and occur at the same modulation frequencies. A suitable measure for the overlap of the $\ket{^1S_0,0,-9/2}$ and $\ket{^1S_0,0,-5/2}$ states in the Floquet mode basis is given by
\begin{equation} \label{eq:state mixture}
    \mathcal{M}_{-9/2,-5/2} := 2\sum_n \vert\braket{^1S_0,0,-9/2\vert n}\braket{n\vert ^1S_0,0,-5/2}\vert^2\;.
\end{equation}
The state mixture measure $\mathcal{M}$ ranges from~0~to~1. It reaches~1 only if exactly two basis states are a 50\% mixture of the target states. Furthermore, $\mathcal{M}$ is zero if there is no state in the basis that mixes the target states.

Thus, the analysis of the eigenstates of $U(T)$ reveals the global dynamics of the full time evolution. Within one period, the time evolution is governed by the stroboscopic evolution operator of Floquet theory (see Appendix~\ref{app:Floquet}), which only leads to minor oscillations of the excited state occupation. Therefore, the effective hyperfine coupling is also small. This means that the $m_I$ are sufficiently good quantum numbers during the stroboscopic time evolution of the system, and indeed, the amplitude modulation is responsible for the spin flipping. When this approximation starts to fail and the stroboscopic evolution is not constant, then additional oscillatory behavior can be observed within one period of the amplitude modulation; e.g. for the transitions at $B=300\;\mathrm{G}$ and $\Omega_\mathrm{E}/2\pi = 40\;\mathrm{MHz}$ in Figure~\ref{fig:rabi close up}. 

The one-period time evolution operator can thus be used to obtain an effective Hamiltonian, i.e. the Floquet Hamiltonian, and an effective time-independent dynamics of the system. However, a perturbative approach to calculate the one-period time evolution operator is unsuitable, since the interaction strength of the time-dependent modulation and the transformed energies of the system in RWA are of the same order of magnitude. Therefore, obtaining $U(T)$ requires a numerical approach.

\section{Stability analysis}\label{sec:stability}
We now investigate the stability of the proposed ONER protocol for our choice of `qubit' transitions between $m_I=-9/2$ and $m_I=-5/2$ in the electronic ground state. The stability is checked with respect to the laser parameters $\theta$, $\Delta$ and $\Omega_\mathrm{E}$. Also, the robustness against variations in the magnetic field $B$ and the amplitude modulation period $T$ is tested. Moreover, we discuss the problem of destructive scattering and atom loss to other states within ONER.

The simulation of the total system reveals that the occupation of the excited state is bounded by $P_{3P1} < 0.01$, in cases of reasonably stable transitions with negligible stroboscopic behavior. The decay rate $\Gamma$ for the $^3P_1$ state of $^{87}$Sr leads to 
\begin{equation}
    N_{\mathrm{sc}} = \frac{P_{3P1} \; \Gamma}{\Omega_\mathrm{N}} < 0.008
\end{equation}
as the total number of spontaneously scattered photons per Rabi cycle, assuming a nuclear Rabi frequency of $\Omega_\mathrm{N}/2\pi > 10\;\mathrm{kHz}$. Therefore, we can neglect heating and atom loss in this case.

\begin{figure*}[!ht]
    \centering
    \begin{subfigure}[b]{0.49\textwidth}
        \centering
        \includegraphics[width=\textwidth]{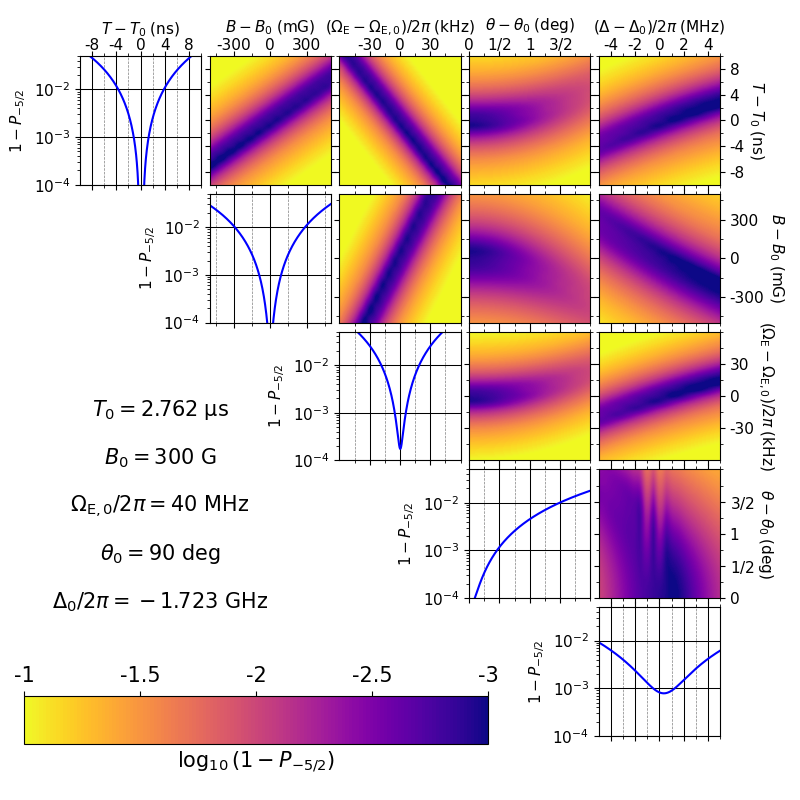}
        \put(-252,240){\small (a)}
        \phantomcaption\label{fig:stability 300 40}
    \end{subfigure}
    \begin{subfigure}[b]{0.49\textwidth}
        \centering
        \includegraphics[width=\textwidth]{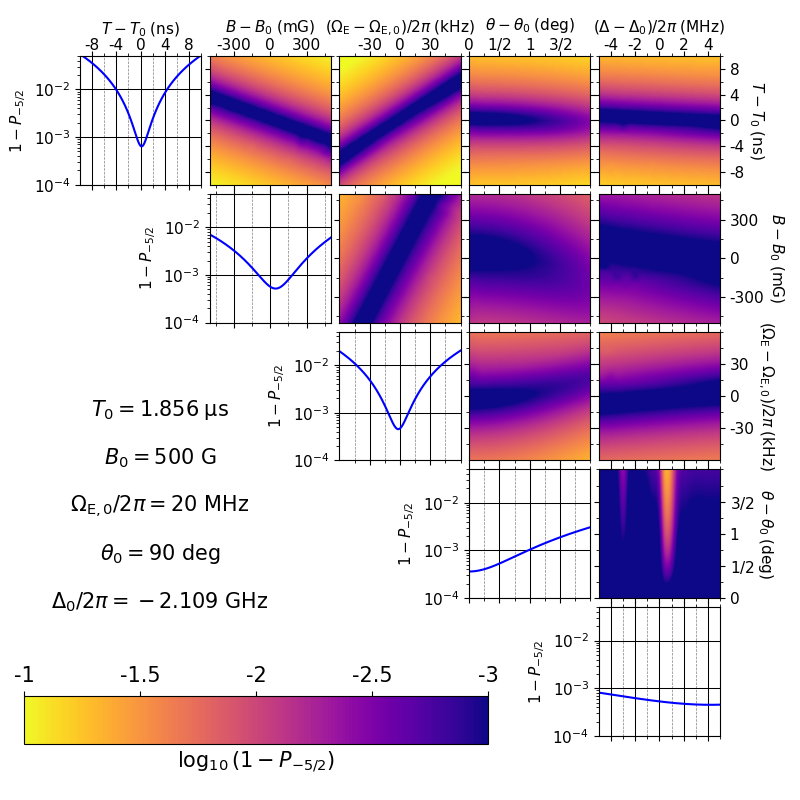}
        \put(-252,240){\small (b)}
        \phantomcaption\label{fig:stability 500 20}
    \end{subfigure}
    \caption{$\pi$-pulse fidelity for variations of the tuning parameters (period of amplitude modulation $T$, magnetic field $B$, electronic Rabi frequency $\Omega_\mathrm{E}/2\pi$, polarization angle $\theta$ and laser detuning from the central transition line $\Delta$) around a spin flip for two different parameter sets, as indicated on the left of each plot. We compute the logarithm of the flip probability $P_{-5/2} = \max_{t\in\tau} \vert\braket{^1S_0,0,-5/2 \vert \psi(t)}\vert^2$ within the simulation time interval $\tau = [0, 50\;\mathrm{\mu s}]$. Diagonal plots illustrate the dependence of the flip probability under variation of one parameter while keeping all other parameters constant at the reference value. Off-diagonal plots illustrate the dependence of the flip probability upon variation of two parameters. Our results demonstrate that fidelities exceeding the fault-tolerant quantum computing threshold of 99.9\% (corresponding to $\log_{10}(1 - P_{-5/2}) = -3$) can be reliably achieved, even if typical experimental noise is taken into account. (a) Scatter matrix for the third peak of the $B=300\;\mathrm{G}$ and $\Omega_\mathrm{E}/2\pi = 40\;\mathrm{MHz}$ transitions. (b) Scatter matrix for the first peak of the $B=500\;\mathrm{G}$ and $\Omega_\mathrm{E}/2\pi = 20\;\mathrm{MHz}$ transitions.}\label{fig:stability}
\end{figure*}

To evaluate the stability with respect to the laser parameters and the magnetic field, we perform simulations with perturbed parameters and investigate the transition probability of the respective target states. We simulate the time evolution and calculate the $\pi$-pulse fidelity of the $\ket{^1S_0,0,-9/2}\rightarrow\ket{^1S_0,0,-5/2}$ transition under perturbed parameters. We apply perturbations individually to each parameter and to combinations of two parameters. The simulation results are graphically summarized in Figure~\ref{fig:stability}. 
We select specific frequencies for the amplitude modulation that are reasonable for experimental verification. For $B=300\;\mathrm{G}$ and $\Omega_\mathrm{E}/2\pi = 40\;\mathrm{MHz}$ we choose the third peak for the amplitude modulation period (see Figure~\ref{fig:stability 300 40}). Additionally, we discuss the stability of the first peak from the $B=500\;\mathrm{G}$ and $\Omega_\mathrm{E}/2\pi = 20\;\mathrm{MHz}$ transitions (see Figure~\ref{fig:stability 500 20}). We refer to these cases as case (a) and case (b).

Noise in the magnetic field up to 0.05\% is tolerable for a $\pi$-pulse fidelity exceeding 99\%, corresponding to 150~mG and 300~mG for case (a) and case (b), respectively. These tolerances are readily achieved with commercial power supplies, $\sim$0.009\%, and are far exceeded with modern stabilization techniques, achieving stability up to $\sim$0.00023\%.~\cite{Borowski2023bfield_stab} Similar fidelities can be achieved with detunings from the optimal laser frequency of up to 4~MHz, a polarization angle perturbation of more than 1 degree, and perturbations in the amplitude modulation period of up to 3~ns ($\sim$0.3\%), for both cases. Noise in the electronic Rabi frequency is tolerable up to 15~kHz $(\sim0.035\%)$ for case (a) and 30~kHz ($\sim0.15\%$) for case~(b). The tolerable noise in the Rabi frequency can be translated into intensity noise by $\delta\Omega_\mathrm{E} / \Omega_\mathrm{E} = \frac{1}{2}\delta \mathcal{I} / \mathcal{I}$, using the relation $\Omega_\mathrm{E} \sim \sqrt{\mathcal{I}}$. Commercial stabilization systems readily achieve an intensity stability of 0.05\%, corresponding to 0.025\% in electronic Rabi frequency. For small variations and fine-tuned control parameters, also fidelities exceeding 99.9\% are possible.
Thus, we observe stable results in both cases with respect to magnetic field perturbations, detuning, polarization angle, amplitude modulation period, and intensity for experimentally reasonable parameter values and variations.

The stability analysis, displayed in Figure~\ref{fig:stability}, shows the dependence of the tuning parameters and the stability with respect to quasi-static variations. The analysis remains valid if the timescale of the variations is large compared to the timescale of the gate operation. If this is not the case, the time-dependence of the variations has to be described with a suitable time-dependent noise model, fitted to the experimentally observable variations of the individual parameters. 
\hl{As shown in Appendix~\ref{app:noise}, measured magnetic-field, laser-frequency, and modulation-source noise in neutral-atom platforms is strongly suppressed at the relevant frequencies $\Omega_{\mathrm N}/2\pi=10$--$100$~kHz, resulting in estimated relaxation rates $\Gamma_{1}\!\ll\!10^{-1}\,\mathrm{s^{-1}}$ ($<10^{-5}$ error per gate). Thus, on the timescale of a single $\pi$-operation, dynamical noise is negligible, and the observed gate stability is limited by quasi-static, shot-to-shot parameter variations quantified by the scatter-matrix analysis. For long-term operation, however, slow drifts or correlated noise could give rise to additional decay channels; in this regime, active stabilization or continuous dynamical-decoupling techniques~\cite{Cai_2012, Mishra2014} may be employed to further suppress decoherence.}

\section{Summary}
In summary, we demonstrated that optical nuclear electric resonance (ONER) is a powerful method for precise and rapid nuclear spin manipulation in a Paschen-Back-like regime. We showed how nuclear spin transitions can be induced by a modulation of the electric field gradient, generated by a periodic change of the electronic environment using a laser field. This technique provides a compelling alternative to methods based on radio frequencies and standard two-photon Raman transitions.

\hl{We focused our analysis on $^{87}$Sr, which features a nuclear spin that couples only weakly to the environment and therefore serves as an excellent long-lived quantum memory~\cite{daley_quantum_2011}. A current limitation in this platform is the achievable nuclear Rabi frequency, typically of order $1~\mathrm{kHz}$~\cite{barnes_assembly_2022,ahmed_coherent_2025}. We showed that the ONER approach can enable single-qubit gate operations with Rabi frequencies exceeding $10~\mathrm{kHz}$, without compromising coherence or scalability. Since these operation rates surpass the thresholds relevant for fault-tolerant quantum computing~\cite{fowler_surface_2012}, the proposed scheme offers a clear advantage for scalable neutral-atom architectures. This demonstrates that advanced optical control in moderate-to-strong magnetic fields can substantially accelerate single-qubit operations in nuclear-spin-based platforms.}

We note that, while this work primarily focused on two nuclear spin states, $^{87}$Sr actually offers 10 different nuclear spin states. We are confident that ONER can be used to encode information in more than two states. This capability should allow for the realization of qudits~\cite{omanakuttan_qudit_2023, zache_fermion_2023, Deutsch2021qudit, ahmed_coherent_2025}, which are highly sought after in quantum computing due to their potential to significantly reduce circuit complexity~\cite{luo_universal_2014} and enhance the performance of algorithms~\cite{campbell_enhanced_2014, chi_programmable_2022, ringbauer_universal_2022}.

As discussed, higher magnetic fields require higher laser intensities, but can also lead to more robust Rabi oscillations. However, proof-of-principle demonstrations in $^{87}$Sr are feasible at magnetic field strengths and laser intensities which are readily accessible in typical laboratories working with strontium. Further improvements of the ONER method, supported by experimental studies, will expand its applicability to other atomic and molecular systems. These efforts may lead to significant advances in nuclear spin control, enhancing the functionality of atomic and molecular quantum devices and unlocking new possibilities for quantum technologies.

\bigskip
\begin{appendix}

\section{Energy correction in the quadrupole Hamiltonian}
\label{app:quadrupole energy}
Here, we discuss the energy correction in the quadrupole Hamiltonian in more detail. We start from the total quadrupole Hamiltonian given in Equation~\eqref{eq:HbHq} and assume that the quadrupole interaction is small relative to the Zeeman interaction. We then apply first-order perturbation theory to obtain the corrected energies as well as the corrected transition energies for both one- and two-level transitions. This leads to
\begin{align}
\begin{split}
    \mathcal{E}_{m_I}^{(1)} 
    &= \b{m_I} H \k{m_I} \\
    &= -\gamma_n B_0 m_I + \left(\frac{3m_I^2}{2} - \frac{I(I+1)}{2} \right) Q_{zz},
\end{split}
\end{align}
yielding the corrected transition energies
\begin{align} \label{eq:trans_energy_corr}
\begin{split}
\Delta \mathcal{E}(m_I-1 \rightarrow m_I) &= -\gamma_n B_0 + \frac{3}{2}(2m_I - 1)Q_{zz} \\
\Delta \mathcal{E}(m_I-2 \rightarrow m_I) &= -2\gamma_n B_0 + \frac{3}{2}(4m_I - 4)Q_{zz}. 
\end{split}
\end{align}
Note that this correction opens the possibility to address specific transitions individually, which is not possible in the case of equidistant Zeeman-splitting alone. These corrected transition energies become relevant when choosing a suitable driving frequency for the nuclear spin system.

\section{Theoretical background of optical nuclear electric resonance}\label{app:ONER}
Following the derivation outlined in Ref.~\citenum{krondorfer_nuclear_2023}, we consider a Hamiltonian of the form
\begin{align}
\begin{split}
    H(t) 
    &= H_\mathrm{E}(t) \otimes \mathbb{1} + \mathbb{1} \otimes H_B + \sum_{\mu\nu} \hat{Q}_{\mu\nu} \otimes \hat{I}_\mu \hat{I}_\nu,
\end{split}
\end{align}
where all non-nuclear-spin related contributions to the Hamiltonian are collected in a `molecular' or 'electronic' part $H_\mathrm{E}(t)$, i.e. all purely electronic interactions in an atom, including interactions with external fields. The interaction of the nuclear spin with an external magnetic field is described by $H_B$, and the last term describes the coupling of electronic degrees of freedom and nuclear spin $\hat{\bm{I}}$ mitigated by the nuclear quadrupole interaction (NQI) tensor $\hat{Q}_{\mu\nu}$. If the Born approximation
\begin{equation}
    \rho(t) = \rho_\mathrm{E}(t)\otimes\rho_\mathrm{N}(t)
\end{equation}
holds for the electronic and nuclear spin density matrix, then the total von Neumann equation $\rmi\partial_t\rho=\com{H(t)}{\rho}$ can be reduced to two effective dynamical equations by taking partial traces, yielding
\begin{align} \label{eq:dyn_mol_efg}
\begin{split}
    \rmi\partial_t\rho_\mathrm{E} &= \com{H_\mathrm{E}(t)}{\rho_\mathrm{E}} + \mathcal{O}\left(\norm{\hat{Q}}\right)\;, \\
    \rmi\partial_t\rho_\mathrm{N} &= \com{H_B + \L \hat{Q} \R_{\mu\nu} (t) \hat{I}_\mu \hat{I}_\nu}{\rho_\mathrm{N}}, 
\end{split}
\end{align}
with
\begin{align} \label{eq:dyn_mol_efg_2}
\begin{split}
    \L \hat{Q} \R_{\mu\nu} (t) &= \tr_{\mathrm{E}}{\left\{\rho_\mathrm{E}(t) \hat{Q}_{\mu\nu}\right\}} + \mathcal{O}\left(\norm{\hat{Q}}^2\right).
\end{split}
\end{align}
Both equations can be solved via pure states, i.e., via a Schrödinger equation instead of the von Neumann ansatz for density operators. As shown in Ref.~\citenum{krondorfer_nuclear_2023} this allows for a simplified analysis of the reduced system, and yields the following guidelines for the choice of suitable laser parameters:
\begin{itemize}
    \item The laser frequency should be chosen such that it matches the transition energy of the ground and excited states of the 'electronic' system with suitable detuning.
    \item The amplitude modulation should be chosen such that it matches the transition energy of the effective spin system (which is mainly given by the Zeeman splitting if the quadrupole interaction can be treated perturbatively).
    \item The amplitude modulation has to result in an adiabatically modulated occupation of the excited state.
\end{itemize}

In the case of strong hyperfine interaction, the Born approximation fails, and the simplified analysis is no longer applicable to obtain an effective Hamiltonian for the target spin system. Nevertheless, the simplified approach delivers a reasonable heuristic that can also deepen the understanding of the more complicated behavior of the singlet-triplet transitions. An effective Hamiltonian can be obtained numerically by employing Floquet theory, as discussed in Section~\ref{sec:floquet}.

\section{Conversion of Rabi frequency and laser intensity}\label{app: dipole moment}
The laser intensity is an important quantity for the ONER protocol. We have chosen the electronic Rabi frequency $\Omega_\mathrm{E}/2\pi$ of the $^1S_0\rightarrow$$^3P_1$ transition in $^{87}$Sr as a measure of the intensity. To convert electronic Rabi frequency and intensity, the dipole moment of the transition, $\mathcal{D} = \vert\bra{^1S_0}\hat{\bm{d}}\ket{^3P_1}\vert = \frac{0.151}{\sqrt{3}}\;\mathrm{a.u.}$ in each component, is needed, in combination with a Wigner factor of $\frac{1}{\sqrt{3}}$.\cite{cooper2018dipole} This connects laser intensity $\mathcal{I} = \frac{1}{2}\varepsilon_0 c E_0^2$ and Rabi frequency via $\hbar\Omega_\mathrm{E} = -\bra{^1S_0}\hat{\bm{d}}\ket{^3P_1}\cdot\bm{E}_0$, for an electric field amplitude $\bm{E}_0$, and yields
\begin{equation}
    \vert\hbar\Omega_\mathrm{E}\vert = \mathcal{D} \sqrt{\frac{2\mathcal{I}}{\varepsilon_0 c}},
\end{equation}
which results in intensities $\mathcal{I}=1\;\mathrm{\frac{W}{cm^2}}, 4\;\mathrm{\frac{W}{cm^2}}, 10\;\mathrm{\frac{W}{cm^2}}$ for Rabi frequencies $\Omega_\mathrm{E}/2\pi = 20\;\mathrm{MHz}, 40\;\mathrm{MHz}, 60\;\mathrm{MHz}$ respectively.

\section{Simulation Details}\label{app:simulation}
As a first step, the total Hamiltonian is set up according to Equation~\eqref{eq:total time dependent Hamiltonian}. 
As basis states we choose $\ket{n,m_J,m_I}$, with $n\in\{^1S_0,^3P_1\}$, $m_J = 0$ for $^1S_0$ state and $m_J\in\{\pm 1, 0\}$ for $^3P_1$ state and $m_I\in\{-9/2,-7/2,...,9/2\}$ as discussed in Section~\ref{sec:87Sr} of the main text. Initial states are chosen within the qubit manifold, i.e.
\begin{equation}
    \ket{\psi_0}\in\left\{ \ket{^1S_0,0,-9/2}, \ket{^1S_0,0,-5/2}  \right\}.
\end{equation}

Since the excited $^3P_1$ state decays with rate $\Gamma = 2\pi*7.48\;\mathrm{kHz}$ into the $^1S_0$ ground state, an open quantum system has to be considered with density matrix $\rho$. The time-evolution of the system can then be described by a Born-Markov Master equation~\cite{breuer_opensystem_2007} of the form
\begin{equation} \label{eq:born markov}
    \rmi\hbar\partial_t \rho = \com{H}{\rho} + \rmi\hbar\sum_\alpha k_\alpha \mathcal{L}[c_\alpha]\rho_S,
\end{equation}
with decay strengths $k_\alpha = \Gamma$ and the Lindblad superoperator $\mathcal{L}$ defined by
\begin{equation}
    \mathcal{L}[c]\rho_S = c\rho_S c^\dagger - \frac{1}{2}\left( c^\dagger c \rho_S + \rho_S c^\dagger c \right)
\end{equation}
for a collapse operator $c$. In our case, the collapse operators are given by
\begin{align}
    \begin{split}
        c_0 &= \ket{^1S_0,0}\bra{^3P_1,0} \otimes \mathbb{1}\;, \\
        c_+ &= \ket{^1S_0,0}\bra{^3P_1,+1} \otimes \mathbb{1}\;, \\
        c_- &= \ket{^1S_0,0}\bra{^3P_1,-1} \otimes \mathbb{1}\;,
    \end{split}
\end{align}
and model the spontaneous decay from the excited state to the ground state under the emission of a linearly, right- or left-circularly polarized photon. \hl{Note that noise-induced decay channels are negligibly small, and only quasi-static variations on a shot-to-shot basis are relevant for the fidelity, as discussed in Appendix~\ref{app:noise}.}

We simulate the time evolution given by Equation~\ref{eq:born markov} in a time interval $\tau$ with the Python library QuTiP~\cite{qutip1,qutip2} using the \textit{mesolve} function. From the simulation results, we extract the occupation of the respective states.

\section{Floquet Theory}\label{app:Floquet}
If a Hamiltonian is periodic in time, i.e. $H(t) = H(t+T)$ for some period $T$ (in our case the period of the amplitude modulation), this structure can be used to decompose the time-evolution of the system into a fast stroboscopic time-evolution in between the periods and an evolution for multiples of that period.\cite{shirley_floquet_1965,eckardt_floquet_2017} The analysis is similar to the Bloch state description for Hamiltonians that are periodic in space.

The time evolution operator satisfies the Schrödinger equation
\begin{equation}
    \rmi\partial_t U(t) = H(t) U(t),
\end{equation}
where we set $t_0=0$ for notational simplicity. Note, however, that the choice of $t_0$ is in principle a gauge, although we can neglect this technicality for our purposes.

Using the periodicity of the Hamiltonian and the group structure of the time-evolution operator, one can decompose the time evolution operator $U(t) = S(t) U(T)^{t/T}$, with a time periodic operator $S(t+T) = S(t)$, called stroboscopic evolution operator, and the one-period time-evolution operator $U(T)$.

The eigenstates $\ket{n}$ of the one-period operator are called Floquet modes, which form a time-independent basis of the underlying Hilbert space. These states satisfy
\begin{equation}
    U(T)\ket{n} = e^{-\rmi \varepsilon_n T} \ket{n},
\end{equation}
with quasi-energies $\varepsilon_n$, that are uniquely defined modulo $\frac{2\pi}{T}$. The time-periodic Floquet states are formed by applying the stroboscopic evolution operator on the Floquet modes, i.e. $\ket{\psi_n(t)} = S(t)\ket{n}$. Note that $\ket{\psi_n(kT)} = \ket{n}$ for all integer multiples of the period $T$.

The time-dependent wavefunction can thus be written as
\begin{equation}
    \ket{\Psi(t)} = U(t)\ket{\Psi(0)} = \sum_n c_n e^{-\rmi \varepsilon_n t} \ket{\psi_n(t)},
\end{equation}
with time-independent coefficients $c_n = \braket{n\vert\Psi(0)}$. Note that the stroboscopic evolution determines the behavior within one period, while the one-period time-evolution operator determines the global evolution across several periods. Thus, knowing the Floquet modes and the coefficients $c_n$ determines the shape of the wavefunction for integer multiples of the period, and a decomposition of the wavefunction into a global evolution and a local stroboscopic evolution is possible.

\section{Noise Analysis}\label{app:noise}
\hl{The influence of dynamical non-static noise on gate fidelity within the operation time can be roughly estimated within an effective driven two-level model with states $\ket{0} := \ket{^1S_0,0,-9/2}$ and $\ket{1}:=\ket{^1S_0,0,-5/2}$
\begin{equation}
    H_\mathrm{eff} = \Omega_\mathrm{N}\sigma_x + \Delta_\mathrm{eff}\sigma_z\,,
\end{equation}
where $\sigma_{x,z}$ the respective Pauli matrices, and $\Omega_\mathrm{N}$ is the nuclear Rabi frequency, and $\Delta_\mathrm{eff}$ an effective detuning. On the timescale of a $\pi$-pulse operation, $\tau_\pi=\pi/\Omega_{\mathrm N}=10{-}100~\mu$s, slow fluctuations in the control parameters $x\!\in\!\{B,\Delta,\Omega_{\mathrm E},\theta,T\}$ are effectively constant and already captured by the scatter-matrix analysis in Section~\ref{sec:stability}. Only fluctuations with spectral weight near the dressed-state splitting, i.e. the nuclear Rabi frequency $\Omega_{\mathrm N}$, lead to irreversible decay of the Rabi oscillations.}

\hl{For weak, stationary noise, the (Floquet-)Bloch--Redfield or filter-function formalisms up to first order yield~\cite{breuer_opensystem_2007, suter_spins_2008, Ithier2005, Bylander2011, Mori2022}
\begin{equation}
    \Gamma_{1}(x)\approx(\partial_xA)^2 S_x(\Omega_{\mathrm N}), 
\label{eq:app_gamma}
\end{equation}
where $S_x(\omega)$ is the one-sided power spectral density of the fluctuating parameter $x(t)$ and $A$ the corresponding control-dependent quantity (e.g.\ effective detuning or nuclear Rabi frequency). Equation~\eqref{eq:app_gamma} describes the noise-induced transverse relaxation.\cite{Ithier2005}
The derivative of the effective detuning can be estimated by the two-level model, with maximal excited state occupation $P = \frac{\Omega_\mathrm{N}^2}{\Omega_\mathrm{N}^2 + \Delta_\mathrm{eff}^2}$, and the stability matrix in Figure~\ref{fig:stability}, which contains the maximal occupation. Using $\Delta_\mathrm{eff}\approx \frac{\partial \Delta_\mathrm{eff}}{\partial x}\delta x$ one obtains
\begin{equation*}
    \left(\frac{\partial\Delta_\mathrm{eff}}{\partial x}\right)^{\!\!2} \approx \frac{1-P_{\delta x}}{P_{\delta x}} \frac{\Omega_\mathrm{N}^2}{(\delta x)^2}\,,
\end{equation*}
with $P_{\delta x}$ the maximal fidelity at parameter variation $\delta x$, which can be read off in Figure~\ref{fig:stability}.}

\hl{Measured technical spectra in optical-clock and tweezer systems show that dynamical noise at $\Omega_{\mathrm N}/2\pi=10$--$100$~kHz is strongly suppressed for all relevant control parameters. Magnetic-field noise acts as detuning noise, transverse in the dressed basis. Active stabilization suppresses spectral components above a few kilohertz, leaving $S_B(\Omega_{\mathrm N})$ below detection limits~\cite{Xu_2019, Borowski2023bfield_stab}, and thus giving a negligible contribution to the decay.}

\hl{Likewise, laser-frequency noise enters as detuning noise. Cavity-stabilized lasers used for Sr and Yb clocks exhibit sub-Hz linewidths and high-offset phase noise far below $10^{-17}\,\mathrm{Hz^{-1}}$ for offsets $\gtrsim1$~kHz~\cite{Davila_Rodriguez_2017}. Note that the phase noise value needs to be multiplied by $\Omega_\mathrm{N}^2$ to yield $S_{\omega}(\Omega_\mathrm{N})$.\cite{Rubiola2010} This leads to negligible decay $\Gamma_1(\omega) \le 10^{-15}\;\mathrm{s}^{-1}$ as $S_{\omega}(\Omega_\mathrm{N}) = \Omega_\mathrm{N}^2 10^{-17} \,\mathrm{Hz^{-1}} \approx 10^{-4}\;\mathrm{s}^-1$, and an estimated slope $(\partial\Delta_\mathrm{eff}/\partial \omega)^2 \approx 10^{-6}$.}

\hl{Fluctuations of the modulation period correspond to frequency noise of the amplitude modulation, translating into a detuning in an effective model and thus act transversely. Phase-noise spectra of low-noise RF sources and AOM drivers are below $10^{-7} \,\mathrm{Hz^{-1}}$ at 1-100~kHz~\cite{Weng2025_OE_AOMphnoise}, leading to $S_{1/T}(\Omega_{\mathrm N}) \approx \Omega_\mathrm{N}^2 10^{-7}\;\mathrm{Hz}^{-1}$. With $(\partial\Delta_\mathrm{eff}/\partial \omega)^2 \approx 10^{-2}$, this yields negligible decay $\Gamma_1(1/T) \le 10^{-1}\;\mathrm{s}^{-1}$.}

\hl{Finally, amplitude (intensity) noise may couple longitudinally as well as transversely. The longitudinal contribution is accounted for by slow phase accumulation within the scatter matrix. The transversal contribution at $\Omega_{\mathrm N}$ is set by the relative intensity noise (RIN), where $S_{\Omega_{\mathrm E}}/\Omega_{\mathrm E}^2=\mathrm{RIN}/4$, since $\Omega_\mathrm{E} \sim \sqrt{\mathcal{I}}$. Stabilized AOM/EOM systems reach $\mathrm{RIN}(\Omega_{\mathrm N})\lesssim10^{-12}\;\mathrm{Hz^{-1}}$~\cite{Wang2020, Arar2017}, leading to $S_{\Omega_{\mathrm E}}(\Omega_\mathrm{N}) \approx 10^{-4}\;\mathrm{s}^{-1}$. Assuming locally an approximately linear dependence $\Omega_\mathrm{N}\sim\Omega_\mathrm{E}$ gives $\vert\partial \Omega_\mathrm{N} / \partial \Omega_\mathrm{E}\vert \ll 1$, due to the different orders of magnitude. Overall, the decay is then bounded by $\Gamma_1(\Omega_\mathrm{E}) \ll 10^{-4}\;\mathrm{s}^{-1}$. Polarization-angle fluctuations are dominated by slow thermal and mechanical drifts and therefore produce only quasi-static noise already included in the scatter-matrix analysis.}

\hl{These estimates are consistent with coherence times of tens of seconds observed in Sr and Yb nuclear-spin qubits, where residual decay is attributed primarily to optical-trap laser noise~\cite{barnes_assembly_2022, ma_universal_2022, jenkins_ytterbium_2022}. Hence, dynamical noise contributes negligibly to gate infidelity: fidelity is limited by quasi-static, shot-to-shot parameter variations that can be described by a scatter matrix, as discussed in Section~\ref{sec:stability}. For the short-term stability and fidelity, noise-induced decay remains negligible.}

\hl{However, regarding long-time behavior and long-time stability, these noise sources may open additional decay channels that can become relevant. In this case, additional stabilization techniques, such as dynamical decoupling or concatenated continuous decoupling~\cite{Cai_2012, Mishra2014}, might be suitable remedies.}
\end{appendix}

\begin{acknowledgments}
We thank R. v. Bijnen and A. Kruckenhauser from planqc, as well as J. Geiger, V. Klüsener, S.L. Kristensen, and F. Spriestersbach from MPQ for stimulating discussions and helpful comments. J.K. Krondorfer, M.Diez and A.W. Hauser acknowledge funding by the Austrian Science Fund (FWF) [10.55776/P36903]. We further thank the IT Services (ZID) of the Graz University of Technology for providing high-performance computing resources and technical support.
S. Pucher and S. Blatt acknowledge funding by the Munich Quantum Valley initiative as part of the High-Tech Agenda Plus of the Bavarian State Government and by the BMBF through the program ``Quantum technologies -- from basic research to market'' (Grant No. 13N16357).
\end{acknowledgments}

\input{output.bbl}

\end{document}

%% file: output.bbl
%